\documentclass[aps,prl,a4paper,showpacs,twocolumn]{revtex4}

\input epsf.tex

\usepackage{amssymb}
\usepackage[latin1]{inputenc}
\usepackage{amsmath}
\usepackage{amsfonts}
\usepackage{bm}
\usepackage{graphicx}
\usepackage{epsfig}
\usepackage{subfigure}
\usepackage{setspace}

\newcommand{\be}{\begin{equation}}
\newcommand{\ee}{\end{equation}}
\newcommand{\ba}{\begin{eqnarray}}
\newcommand{\ea}{\end{eqnarray}}

\newcommand{\ket}[1]{\mbox{$ | #1 \rangle $}}

\begin{document}

\title{Fidelity of an optical memory based on stimulated photon echoes} \date{\today} \pacs{03.65.Yz; 42.25.Hz; 42.50.Md;}
\author{M. U. Staudt$^{1}$, S. R. Hastings-Simon$^{1}$, M. Nilsson$^{1}$, M. Afzelius$^{1}$,
V. Scarani$^{1}$, R. Ricken$^{2}$, H. Suche$^{2}$, W.
Sohler$^{2}$}

\author{W. Tittel$^{1}$}
\altaffiliation{Now at: Institute for Quantum Information Science,
University of Calgary, Canada}
\author{N. Gisin$^{1}$}
\affiliation{$^{1}$ Group of Applied Physics, University of
Geneva, CH-Geneva, Switzerland\\ $^{2}$ Angewandte Physik,
University of Paderborn, 33095 Paderborn, Germany }

\begin{abstract}
We investigated the preservation of information encoded into the
relative phase and amplitudes of optical pulses during storage and
retrieval in an optical memory based on stimulated photon echo. By
interfering photon echoes produced in a Ti-indiffused single-mode
Er-doped LiNbO$_{3}$ waveguiding structure at telecom wavelength,
we found that decoherence in the atomic medium translates only as
losses (and not as degradation) of information, as long as the
data pulse series is short compared to the atomic decoherence
time. The experimentally measured value of the visibility for
interfering echoes is close to 100 $\%$. In addition to the
expected three-pulse photon-echo interferences we also observed
interference due to a four-pulse photon echo. Our findings are of
particular interest for future long-distance quantum communication
protocols, which rely on the reversible transfer of quantum states
between light and atoms with high fidelity.
\end{abstract}

\maketitle

\begin{center}
\begin{figure}
\epsfxsize=0.4\textwidth \epsfbox{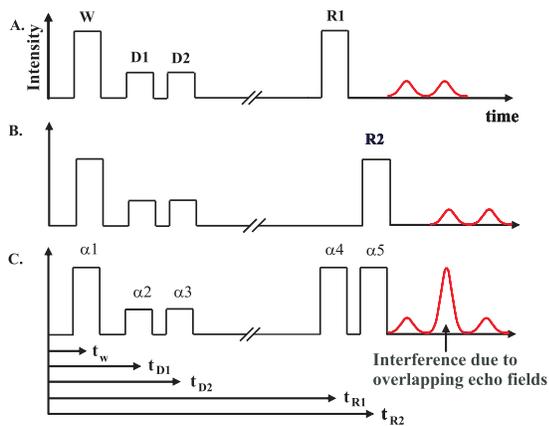} \caption{Illustration
of the sequence of pulses for the interference of photon echoes.
The data is read out twice and the phase between data and read
pulses is changed to produce interference in the central time bin
(see text for details).}\label{fig1}
\end{figure}
\end{center}

\begin{center}
\begin{figure}
\epsfxsize=0.4\textwidth \epsfbox{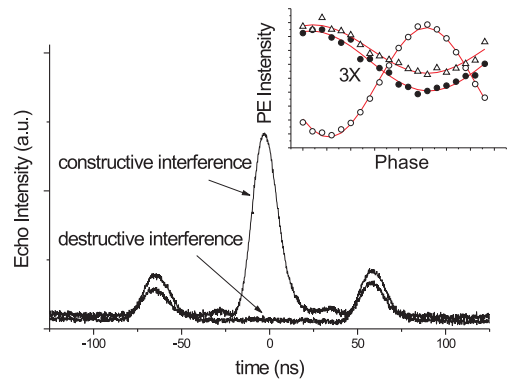}
 \caption{Photon echo signals for constructive and destructive interference. Inset:  Central echo peak
intensity $\circ$, left $\bullet$ and right $\vartriangle$ side
peak intensity  as a function of phase. The phase modulation of
both side peaks is synchronized and is opposite to the principle
peak (Note that the amplitude of the side peaks was multiplied
with a factor of 3). This is due to an interference with
four-pulse photon echo (4PE) as described in more detail in the
text.}\label{fig2}
\end{figure}
\end{center}

\begin{center}
\begin{figure}
 \epsfxsize=0.4\textwidth \epsfbox{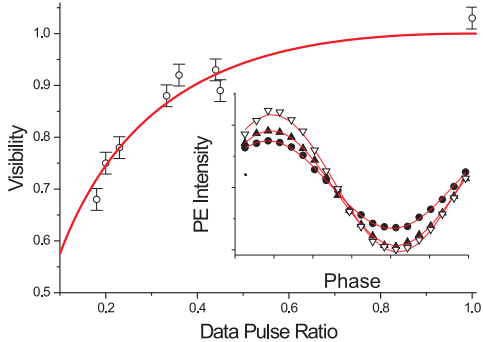}
\caption{The visibility as a function of the ratio between the two
time-bin pulses is shown. Experimental points $\circ$ are in good
agreement with Eq.~(\ref{visi}), which contains no free parameter.
Inset: The area under the interfering photon echoes is plotted as
a function of the phase for different incoming time-bin amplitude
ratios ($\bullet$ V=0.68, $\blacktriangle$ V=0.93, $\triangledown$
V=1.04). The interference visibility (V) is extracted from a
sinusoidal fit. }\label{fig3}
\end{figure}
\end{center}

\begin{center}
\begin{figure}
\epsfxsize=0.4\textwidth \epsfbox{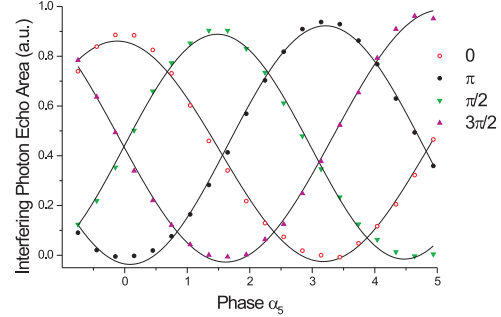} \caption{Area under
the echo as function of the phase of the second read-out pulse.
All four classical states, analogous to the quantum states used in
four-state quantum cryptography protocols, are stored, retrieved
and analyzed with close to 100 \% fidelity. This is possible even
though the probability of retrieval from the memory is only a few
percent, limited by the efficiency of the photon echo process and
by decoherence processes in the storage material.}\label{fig4}
\end{figure}
\end{center}

Transfer of coherence properties between light and atoms can be
investigated through interferometric and spectroscopic techniques.
These studies are of fundamental interest, but also deliver
important information for
future applications in the field of quantum information science.\\
In quantum communication schemes, such as quantum cryptography,
non-orthogonal states of light are used as information carrier.
The encoding of information into the relative phase and amplitudes
of a time-bin qubit has proven to be well suited for transmission
over long distances, because this coding is robust to the
decoherence mechanism in optical fibers \cite{brendel99,gisin02}.
However, the extension of quantum communication to arbitrary
distances relies on the availability of {\em quantum memories},
which are key to the building of a quantum repeater
\cite{Briegel98}. Although significant progress has recently been
reported \cite{julsgaard04,chaneliere05,eisaman05}, coherent,
reversible transfer of quantum information from photons to atoms
with high fidelity and efficiency remains an important
and open challenge.\\
From this perspective, it is important to understand {\em how the
fidelity of a time-bin qubit evolves when the information is
stored in a quantum memory}. It is the primary objectives of this
Letter to address this issue. In particular, we show a case in
which the decoherence in the atomic medium is a state-independent
coupling with the environment: its effect on the retrieved signal
is therefore only {\em losses}, i.e. a decrease of the retrieval
probability. By post-selecting only cases when photons are
actually emitted, one retrieves uncorrupted information, which
does not require complicated classical or quantum error
correction.
\\We work in the framework of a recent proposal \cite{kraus05}, and of first experimental studies
\cite{alex06}, for storage of time-bin qubits based on controlled,
reversible, inhomogeneous broadening. This is a {\em photon-echo
type approach} to quantum memories with a theoretical efficiency
of 100~$\%$. Photon-echoes are well known for storage of classical
optical pulses \cite{moss82,mitsunaga92a} as well as for being a
phase-preserving process. However, the storage of information
encoded in the amplitude and relative phase of subsequent optical
pulses, crucial for the proposal under study, has received only
limited attention so far \cite{arend93,elman96}.
\\The photon echo
experiments reported here were done using an Er$^{3+}$ doped
LiNbO$_{3}$ crystal with a waveguiding structure. To our
knowledge, these are the first reported photon echo experiments in
Er$^{3+}$:LiNbO$_{3}$ waveguides, LiNbO$_{3}$ being widely used as
a non linear material in integrated optics. In the experimental
set-up the light is guided entirely through standard
telecommunication fibers, integrated intensity/phase modulators,
polarization controllers and the Ti-indiffused
Er$^{3+}$:LiNbO$_{3}$ waveguide in a mono-mode structure, thus the
assumption of only one dimension, often used in theoretical
calculations, is fully justified.\\A common approach to storage
and retrieval of light using photon echoes is based on {\em
three-pulse photon echo (3PE)}, also known as stimulated photon
echo \cite{mitsunaga92a}. In this process a first strong optical
"write" pulse prepares the medium. The "data" pulses, a sequence
of pulses encoding the information to be stored, are sent into the
medium some time after the write pulse. In order to retrieve the
information, a third strong "read pulse" is used, which causes a
photon echo to be emitted afterwards. If certain conditions for
excitation energy and absorption depths are met, the echo is to a
high degree an amplitude and phase replica of the stored data
pulses. A common physical picture used to explain the 3PE is that
the write pulse creates an atomic coherence and the data pulses
transfer the coherence into a frequency-dependent population
grating in the ground and excited states. The read pulse scatters
off the grating, forming echoes a time after the read pulse, which
is equal to the time separation between write and data
pulse.\\Now, consider a data field consisting of two pulses (D1
and D2) with a amplitude ratio $R$ and phase relation $\varphi$
(see Fig.~\ref{fig1} a.). The 3PEs appears at times
$t_e=t_r+t_{Di}-t_w$ $(i=1,2)$, where $t_r$ is the time of
read-out, $t_{Di}$ the time of data pulse Di $(i=1,2)$ and $t_w$
the arrival time of the write pulse. The echoes will thus be
$dt=t_{D2}-t_{D1}$ apart.
Because the efficiency of the 3PE is at best a few percent
\cite{wang99}, much of the frequency-dependent population grating
is preserved in the atomic ensemble after the read pulse.
Therefore more echoes can be produced by sending in several read
pulses. In our experiment, two subsequent read pulses were used to
produce two copies of the data pulse. If we chose the distance
between the read pulses to be $dt$, the same as the distance
between the two data pulses D1 and D2, the echo of the second data
pulse read out by the first read pulse (D2$|$R1), and the echo of
the first data pulse read out by the second read pulse (D1$|$R2),
will be indistinguishable and thus interfere (Fig.~\ref{fig1} c.).
The phase of a 3PE is controlled by the relative phase of the
write, the data and the read pulse. The write pulse has a phase
$\alpha 1$ , the data pulses a phase $\alpha 2/3$, and the read
pulses a phase $\alpha 4/5$. Thus one can obtain constructive or
destructive interferences by carefully choosing the different
phases of the input pulses \cite{ref}. This is true provided that
the phase or amplitude coherence is not lost partially or totally
during storage and retrieval.\\
The data pulse above is related to what is known in quantum
communication as a time-bin qubit. A {\em time-bin qubit}
\cite{brendel99} is a coherent superposition of a photon being in
two time-bins, separated by a time difference long compared to the
coherence time of the photon. It can be written in the form: \ba
\label{main}
\ket{\psi}=c_0\ket{1,0}+c_1e^{i\varphi}\ket{0,1}\label{qubit}\ea
where $\ket{1,0}$ ($\ket{0,1}$) stands for a photon being in the
first (respectively the second) time-bin and $\varphi=\alpha
2-\alpha 3$ for the relative phase.\\In the present experiment
classical coherent pulses were used (see Fig.~\ref{fig1} c.).
These {\em time-bin pulses} are two coherent pulses such that the
width of each pulse is much smaller than the temporal spacing $dt$
between the two pulses; the state of light is a Poisson
distribution of photons ($n\sim 10^8$), each
of which is in the state described by Eq.~(\ref{qubit}).\\
The output field amplitude of a 3PE, taking $t_w$=0 and assuming
that the whole storage process takes place on a time scale small
compared to the radiative lifetime, will be reduced by a factor of
$e^{-2 t_{Dj}/T_2}$ , where $T_2$ is the \textit{atomic}
decoherence time. Therefore a time-bin qubit absorbed by a photon
echo material will be emitted as follows, assuming that the photon
echo amplitude is linear compared to that of the input field: \ba
\label{decoherence} \ket{\Psi}\sim [e^{-2t_{D1}/T_2} c_0
\ket{1,0}+ e^{-2 t_{D2}/T_2}
c_1 e^{i\varphi} \ket{0,1}]\ket{{\cal E}_0}\nonumber\\
+\lambda \ket{0,0} {{\cal E}_1}\ea here $\ket{{\cal E}_0}$ and
$\ket{{\cal E}_1}$ are the states of the environment to which the
memory couples. The information encoded in the time-bin is
preserved provided $dt=t_{D2}-t_{D1}\ll T_2$ and provided the
process has not modified the pulse in such a way, that the width
of the echo is $\sim dt$. Indeed in this case
Eq.~(\ref{decoherence}) can be simplified to $\ket{\Psi}\sim
e^{-2t_{D1}/T_2}\ket{\psi} \ket{{\cal E}_0}+\lambda
\ket{0,0}\ket{{\cal E}_1}$. It follows that even if atomic
decoherence has acted during a long time ($t_{D1}\sim T_2$) it
does not influence the amplitude ratio or phase difference of the
time-bin information. By means of postselecting the cases where a
detection is obtained one can thus reach a very high fidelity,
however at the expense of a smaller retrieval probability as
compared to simply detecting the vacuum component. A memory having
these characteristics can be, depending on the application,
advantageous compared to one with high retrieval probability and low fidelity.\\
The retrieved time-bin pulses (photon echoes) shown schematically
in Fig.~\ref{fig1} c. interfere constructively or destructively
depending on the phase difference $\varphi$ and the phases of the
read pulses. The visibility V of the interference should only be a
function of the relative amplitudes of the \emph{incoming}
time-bin pulses: \ba V&=& 2\frac{\sqrt R}{1+R}\label{visi},\ea
with ratio $R=c_0^2/c_1^2$.\\Note that one could also describe our
experiment as a setup containing two interferometers, as used for
phase-coding quantum cryptography \cite{gisin02}: One
interferometer prepares the time-bin qubits, i.e. here our two
data pulses, while the second allows the projection measurement,
i.e. our two read pulses.
\\Now we describe the experimental setup, which is similar to the one used in \cite{staudt06}. The output from an external-cavity cw diode laser
(Nettest Tunics Plus) was gated by a combined phase and intensity
modulator and followed by an intensity modulator, both fiber-optic
produced by Avanex. The first modulator created the five
excitation pulses and applied phase shifts to some of the pulses,
depending on the particular experiment, the second modulator was
synchronized to the first one and used to improve the
peak-to-background intensity ratio. The pulses had durations of
$t_{pulse}$=15 ns, with a clock frequency of 30 Hz. The first data
pulse was created at t$_{D1}=0.6~\mu$s and the time between the
data pulses was typically \textit{$dt=60$ns} and the read-out
pulses were delayed with regard to the data pulses by 1 to 2
$\mu$s. The pulses were then amplified by an EDFA (Erbium Doped
Fiber Amplifier). In order to obtain a good background suppression
($>$ 70 dB) and to avoid spectral holeburning by the EDFA, we
placed an additional acousto-optical modulator between the optical
amplifier and the input of the pulse-tube cooler, which opened
only for the series of pulses and suppressed light for all other
times. The light was then coupled into the Er$^{3+}$-doped
LiNbO$_3$ crystal inside the pulse tube cooler (Vericold), where
the crystal was cooled to about 3.4 K. The resulting peak powers
were in the range of 5 mW for the write pulses at the refrigerator
input (and on the order of 1 mW for the other pulses). The photon
echo was detected by a fast detector (1611v, New Focus) after the
pulse-tube cooler.\\The z-cut LiNbO$_3$ was Erbium doped over a
length of $10~mm$ by indiffusion of an evaporated $8~\mu m$ thick
Er-layer at $1130~^0C$ for $150~h$, leading to a Gaussian
concentration profile of $8.2~\mu m$ 1/e penetration depth and
$3.6\times~10^{19}~cm^{-3}$ surface concentration. The guiding
channel was fabricated by indiffusion of a $7~\mu m$ wide, $98~nm$
thick Ti-stripe at $1060~^0C$ for $8.5~h$, leading to a mono-mode
guide with a mode size of $4.5\times~3\mu m$ FWHM intensity
distribution \cite{Baumann97}. The light was injected and
collected with standard optical fibers into a waveguide of a
diameter of $9~\mu m$. A magnetic field  of about 0.2 Tesla was
applied parallel to the C$_{3}$ axis. This reduces decoherence due
to spectral diffusion \cite{Sun02}, resulting in a decoherence
time of about $T_2 \sim 6 \mu s$.\\ Fig.~\ref{fig2} shows typical
interference patterns for constructive and destructive
interference. Here the input data pulses had the same amplitude,
thus R=1. We scanned the phase difference between the two
interfering photon echoes continuously by varying phase $\alpha 2$
using the intensity/phase modulator and obtained a clear
modulation of the photon echo interference signal (see
Fig.~\ref{fig2}).\\To extract the visibility we measured the
background-subtracted area under the echo interference, and
plotted the area as a function of the applied phase. The
background was obtained by fitting the signal on either side of
the side peaks. We have verified by several measurements, that the
detection background was of purely electronic origin and that no
coherent or incoherent background light was interfering with the
echoes.\\The large number of echoes produced by our pulse sequence
(which is reduced to the echoes of interest in Fig.~\ref{fig1} c.)
can easily lead to interference with subsidiary echoes, which has
to be avoided by carefully choosing the time delays between
pulses. Yet, as can be seen in Fig.~\ref{fig2} the side peaks also
show a modulation, which we found to be due to an interference
with higher-order echoes produced by four excitation pulses (4PE)
\cite{ref2}. The 4PE detected is much smaller than the 3PE, which
results into a smaller visibility as compared to pure 3PE
interference (see inset in Fig.~\ref{fig2}). These higher-order
types of echoes have been observed previously and have been
denoted virtual echoes \cite{zac00}.
In order to demonstrate that our PE based measurement setup is
analogous to an interferometer for analyzing the time-bin pulses,
we also performed visibility measurements using time-bin pulses
having different relative amplitudes. As expected, the extracted
visibility increases with amplitude ratio and it follows, within
the experimental error, the theoretical curve calculated using
Eq.~(\ref{visi}). Perfect visibility was reached in the case of
equal amplitudes (see Fig.~\ref{fig3}). Note that the experimental
error is in principal larger for equal time-bin amplitudes, as the
method of background subtraction is more sensitive to noise when
the photon echo signal is small, i.e. at the point of destructive
interference. The error bars for all depicted data points in
Fig.~\ref{fig3} are calculated from standard deviations of a large
number of measurements for $R=1$, setting thus an upper limit.\\In
Fig.~\ref{fig4} the area under the echo is plotted as a function
of the phase $\alpha 5$ that is applied to the second of the
\emph{read} pulses. While this phase is scanned, the phase $\alpha
3$ of the time-bin pulse is kept constant at: 0, $\pi/2$, $\pi$,
and $3\pi /2$ and all other phases are kept at zero. This is
conceptually analogous to preparing four different time-bin qubits
states of two conjugate basis on the equator of the Poincar\'{e}
sphere, as it is widely used in quantum cryptography in the
so-called BB84 \cite{BB84} or four state protocol. While in
quantum cryptography setups the projection measurement is done
with an interferometer \cite{gisin02}, we project the state with
photon echoes using two read pulses. Note that the photon echo
process thus serves two purposes, \emph{storage/retrieval and
analysis} of the state.
Our results show that the relative phase and amplitude ratio of
time-bin pulses can be preserved during storage in the optical
memory. Apart from the variation of the visibility due to the
change of ratio between the two time-bins, no further reduction is
observed, despite the fact that the atomic coherence time is such
that a significant part of the atomic coherence is lost during the
storage time. This can be interpreted in the following way:
external perturbation of the atomic coherence in the Erbium ions
reduces the macroscopic dipole moment, representing a loss of
coherent ions, which reduces the size of the coherent emission.
The ions that have undergone no or small decoherence, however,
still retain the phase and amplitude information of the incoming
excitation fields, which make it possible to store and retrieve
information with high fidelity despite the decoherence in the
photon echo material. This is true as long as the time separations
between the time bins is comparable to or larger than the
decoherence time, as discussed in connection to
Eq.~(\ref{decoherence}). Due to the possibility of post-selection
a nearly perfect fidelity can be obtained being promising for a
future CRIB based quantum memory \cite{kraus05}.

This work was supported by the Swiss NCCR Quantum Photonics and by
the European Commission under the Integrated Project Qubit
Applications (QAP) funded by the IST directorate as Contract
Number 015848. Additionally M.A. acknowledges financial support
from the Swedish Research Council and W. T. by iCORE.



\begin{thebibliography}{99}
\bibitem{brendel99}J. Brendel et al., Phys. Rev. Lett. {\bf82}, 2594, 1999
\bibitem{gisin02}N. Gisin et al. Rev. Mod. Phys. {\bf74}, 145 (2002)
\bibitem{Briegel98}H. J. Briegel et al., Phys. Rev. Lett. {\bf81}, 5932 (1998)
\bibitem{julsgaard04}B. Julsgard et al., Nature {\bf432}, 482 (2004)
\bibitem{chaneliere05}T. Chaneli\`{e}re et al., Nature {\bf438}, 833 (2005)
 \bibitem{eisaman05}M. D. Eisaman et al., Nature {\bf438}, 837 (2005)
\bibitem{kraus05}B. Kraus et al., Phys. Rev. A {\bf73} 020302(R) (2006)
\bibitem{alex06}A. L. Alexander et al., Phys. Rev. Lett. {\bf96}, 043602 (2006)
\bibitem{moss82}T. W. Mossberg, Opt. Lett. {\bf7}, 77 (1982)
\bibitem{mitsunaga92a}M. Mitsunaga, Optical and Quantum Electronics {\bf24}, 1137 (1992)
\bibitem{arend93}M. Arend et al., Opt. Lett. {\bf18}, 1789 (1993)
\bibitem{elman96}U. Elman et al., J. Opt. Soc. Am. B. {\bf13},
1905 (1996)
\bibitem{ref}In the case of a simple 3PE, where the excitation pulses have the
phase $\alpha_{i}$, $i=1,2,3$, the echo will have the phase:
$\alpha_{e}=\alpha_{1}-\alpha_{2}-\alpha_{3}$.
\bibitem{wang99}T. Wang et al., Phys. Rev. A {\bf60} 757(R) (1999)
\bibitem{staudt06}M. U. Staudt et al. Optics Communn. in press
\bibitem{Baumann97}I. Baumann et al., Applied Phys. A {\bf164}, 33 (1997)
\bibitem{Sun02}Y. Sun et al., J. Lumin.  {\bf98},281, 2002
\bibitem{ref2}
The 4PE phase $\Theta$ depends on the contributing pulse phases
for our configuration for the left and right side peak as follows:
$\Theta_1=\alpha_1 -2 \alpha_2 + \alpha_3- \alpha_5+ \pi$ and
$\Theta_2=\alpha_1 -\alpha_2 + \alpha_4-2 \alpha_5+ \pi$. As can
be seen in Fig.~\ref{fig2} the dependency of the phase of the side
peaks is displaced by $\pi$ from the principal peak leading into a
flip of maximum and minimum.
\bibitem{zac00} Z. Cole, master thesis, Montana State University,
Montana, USA, 2000
\bibitem{BB84}C. H. Bennett, G. Brassard, in Proceedings of the IEEE International Conference on Computers,
Systems and Signal Processing, Bangalore, India (IEEE, New
York,1984), p. 175







\end{thebibliography}
\end{document}